# Sparse pixel image sensor

Lukas Mennel[1], Dmitry K. Polyushkin[1], Dohyun Kwak[1], and Thomas Mueller[1*]

[1]Vienna University of Technology, Institute of Photonics, Gußhausstraße 27-29, 1040 Vienna, Austria

*Corresponding author: thomas.mueller@tuwien.ac.at

**As conventional frame-based cameras suffer from high energy consumption and latency, several new types of image sensors have been devised, with some of them exploiting the sparsity of natural images in some transform domains. Instead of sampling the full image, those devices capture only the coefficients of the most relevant spatial frequencies. The number of samples can be even sparser if a signal only needs to be classified rather than being fully reconstructed. Based on this mathematical framework, we developed an image sensor that can be trained to classify optically projected images by reading out the few most relevant pixels. The device is based on a two-dimensional array of metal-semiconductor-metal photodetectors with individually tunable photoresponsivity values. We demonstrate its use for the classification of handwritten digits with an accuracy comparable to that of conventional systems, but with lower delay and energy consumption.**

The Nyquist-Shannon sampling theorem[1] establishes a lower bound for the sampling rate required to capture any given signal of finite bandwidth without loss of information. The original signal can then be perfectly reconstructed from the samples. For many practical signals, however, reconstruction may still be possible from far fewer samples – or measurements – than required by the sampling theorem. This can be understood from the fact that these signals may contain much redundant information or, more precisely, are sparse when represented in a proper domain or basis. Natural images, for example, are known to be sparse in the Fourier or Wavelet domains, which is exploited in several types of transform coding schemes, including the JPEG and MPEG standards[2].

Compressed sensing (CS) is a mathematical framework for the recovery of sparse signals from few measurements[3,4,5,6]. In CS, a signal that is incoherently (e.g. randomly) sampled at the encoder side, can be reconstructed at the decoder by finding the sparsest solution of an underdetermined linear system. Both sampling and compression are performed simultaneously to reduce the number of measurements at the expense of increased computational cost for signal reconstruction.

By combining CS with statistical learning, the number of required measurements can be further reduced, particularly if a given signal only needs to be assigned to one of a few categories, or classes, rather than being fully reconstructed. This can be achieved by using a task-specific basis, learned from data, instead of a generic one such as Fourier or Wavelet. In the sparse sensor placement optimization for classification (SSPOC) algorithm[7,8], the data are not sampled randomly, but a few representative measurement locations are identified from training data. Subsequent samples can then be classified with performance comparable to that obtained by processing the full signal.

Several new types of image sensors have been developed in recent years[9], targeting lower energy consumption and latency than their conventional frame-based counterparts. Many of those devices emulate certain neurobiological functions of the retina, either using complementary metal-oxide-semiconductor (CMOS) technology (silicon retina)[10,11,12] or emerging device concepts[13,14,15,16,17,18]. CS has likewise led to new types of image acquisition systems, such as single-pixel cameras[19], coded aperture imagers[20], and CMOS CS imaging arrays[21,22]. SSPOC, on the other hand, has inspired applications in dynamics and control[23,24], but has to the best of our knowledge not been employed in an imaging device yet. Here, we present a hardware implementation of this algorithm, based on a two-dimensional array of tunable metal-semiconductor-metal (MSM) photodetectors. Each of these detectors can be addressed individually and their photoresponsivity values can be set by the application of a bias voltage. The device is fully reconfigurable and we demonstrate its use for the classification of handwritten digits from the MNIST dataset with an accuracy comparable to that of conventional systems, but with substantially lower delay and energy consumption.

Let us first lay out the operation principle of the image sensor (Figure 1a), exemplified by a simple linear classification problem. We restrict ourselves to binary classification, where an optical image, that is projected onto the chip, is assigned to one of two possible classes. The image is represented by a vector $\mathbf{p} = (P_1, P_2, ..., P_n)^T$ in an $n$-dimensional vector space $\mathbb{R}^n$, where $P_k$ is the optical power at the $k$-th pixel. Unlike in conventional imagers, the photoresponsivity of each pixel is not fixed, but varies over the face of the chip. We aggregate the photoresponsivity values into a vector $\mathbf{r} = (R_1, R_2, \ldots, R_n)^T \in \mathbb{R}^n$, where $R_k$ denotes the responsivity of the $k$-th detector. A linear classifier is a predictor of the form[25]

$$y = \sigma(\mathbf{r}^T \mathbf{p}), \quad (1)$$

where $\sigma$ is a threshold function that maps all values of the inner product $\mathbf{r}^T\mathbf{p}$ below a certain threshold (bias) to the first class and all other values to the second class (Figure 1b). Physically, the inner product is implemented by simply summing up the photocurrents produced by all $n$ detector elements, $I_{\text{tot}} = \sum_{k=1}^n I_k = \sum_{k=1}^n R_k P_k = \mathbf{r}^T\mathbf{p}$. By thresholding $I_{\text{tot}}$, a binary output is obtained that is representative of the two classes. $\mathbf{r}$ is learned from a set of labeled training data. A generalization to multi-class problems can be achieved by splitting pixels into subpixels[13, 26] which allows for a physical implementation of a responsivity matrix $\mathbf{R}$.

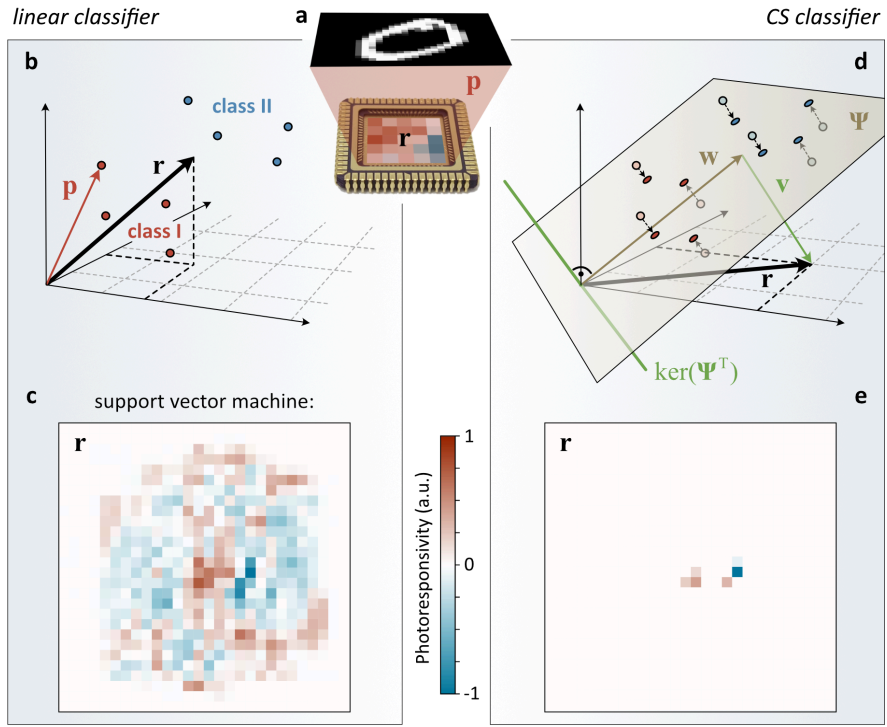

**Figure 1 | Theoretical background and operation principle. a,** Schematic illustration of the setup. An optical image $\mathbf{p}$ is projected onto the face of the image sensor with photoresponsivity values $\mathbf{r}$ that vary from pixel to pixel. **b,** A binary linear classifier assigns an image to one of two possible classes I or II, depending whether or not the inner product $\mathbf{r}^T\mathbf{p}$ is larger than some threshold. In our implementation, the inner product is realized by summing up the photocurrents produced by all detector elements. **c,** Photoresponsivities for a sensor that has been trained as a linear SVM for the classification of zeros and ones from the MNIST dataset. Almost all pixels exhibit non-zero

photoresponsivity values. **d,** Natural images have low-dimensional structure. This allows to construct a sparse photoresponsivity vector **r** for classification. **e,** Results for the same binary classification task as in c. Comparable performance is achieved with 99.2% of the detector elements having zero responsivity.

In Figure 1c we plot **r** for a linear support vector machine (SVM) that is trained to classify handwritten zeros ("0") and ones ("1") from the MNIST dataset. 90% of randomly picked images are used for training and the remaining 10% for assessment. Almost all photodetectors are active, with varying responsivity values, and a classification accuracy of 99.8% is reached.

We now aim to obtain a comparable performance by selecting a small, optimal subset of detectors, or pixels. Figure 1d provides a geometrical interpretation of the algorithm[7,8]. A $d$-dimensional feature space $\Psi$, that spans the $d \ll n$ most significant variations among the training data, is calculated using principal component analysis (PCA)[25]. For categorical decisions, a measurement **p** is projected into this low-dimensional subspace ($\Psi^T: \mathbb{R}^n \to \mathbb{R}^d$) and a linear classifier, described by the weight vector $\mathbf{w} \in \mathbb{R}^d$, is then applied therein: $y = \sigma(\mathbf{w}^T \Psi^T \mathbf{p})$. In image space coordinates, this expression resembles equation (1) with a photoresponsivity vector $\mathbf{r} = \Psi \mathbf{w}$. Note, however, that there exists an infinite number of solutions for **r**, because adding any vector **v** in the null space (kernel) of $\Psi^T$ projects to the very same **w** in feature space. We seek the sparsest solution for **r**, that is the one that has at most $d$ nonzero elements: $\|\mathbf{r}\|_0 \leq d$. As shown by the CS community[3,4,5,6], $\ell_1$-minimization leads to a convex optimization problem that can be efficiently solved with modern methods to find a good approximate solution:

$$\min_{\mathbf{r}} \|\mathbf{r}\|_1, \quad \text{s.t.} \quad \Psi^T \mathbf{r} = \mathbf{w}. \qquad (2)$$

Figure 1e presents the results for the same binary MNIST classification task as before. Here, the data are projected into a six-dimensional PCA subspace ($d = 6$) in which a SVM is trained for classification. The photoresponsivity vector **r** is calculated by $\ell_1$-minimization of (2) using the PySensors package[27] in Python and is plotted in Figure 1e. Although less than 0.8% of the total pixels (6 out of 784) exhibit a responsivity $R \neq 0$, the classifier performs nearly as well as the SVM applied to the full image, and an accuracy of 99.1% is achieved. Importantly, energy consumption and delay are substantially reduced, as both scales linearly with the number of detector elements being read out. We stress that it is not possible to obtain this result by merely thresholding **r** in Figure 1c, as can be seen from Supplementary Figure S1.

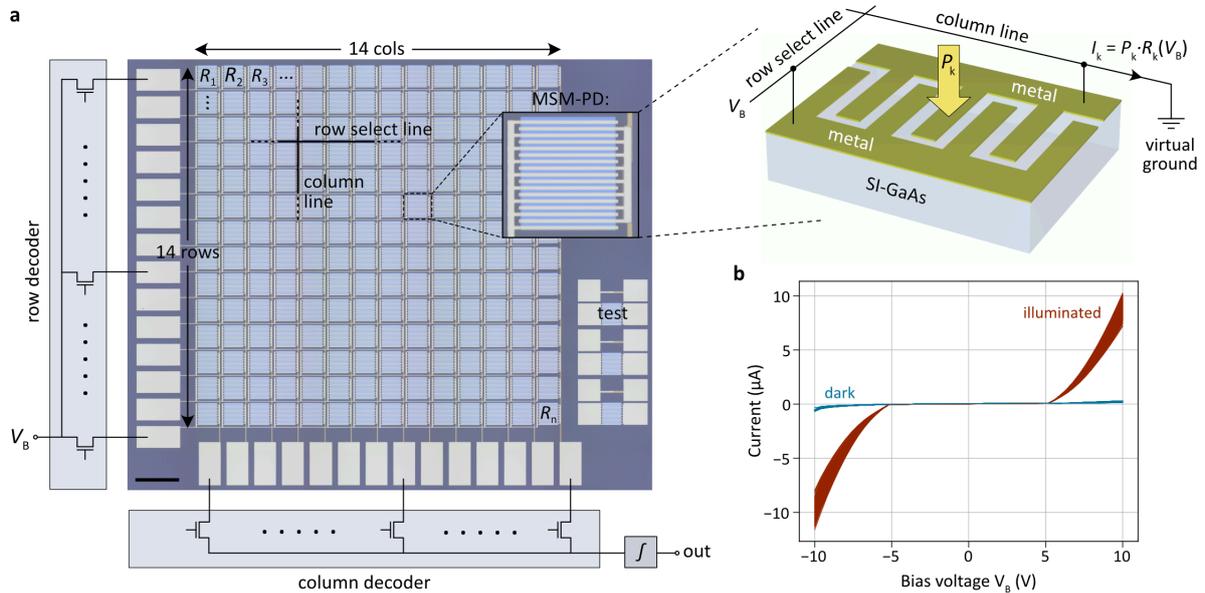

**Figure 2 | Image sensor architecture and characterization. a,** Microscope image of the sensor, with schematic illustrations of the external row/column decoders and integrating output (left). Scale bar, 200 μm. A detailed view of one of the MSM photodetectors is presented in the inset and a schematic illustration is in the picture to the right. **b,** Bias voltage dependent device currents for all 196 detectors with (red lines) and without (blue lines) optical illumination (~160 W/m²). The detectors are operated in the range ±5 V to ±10 V.

In Figure 2a we present the actual device implementation. The sensor is fabricated on a semi-insulating gallium arsenide (SI-GaAs) wafer, with two metal layers for routing of the electrical signals, using standard technology and without high temperature process steps. Details are provided in the Methods section. GaAs is preferred over silicon (Si) because of its shorter absorption and diffusion lengths, which both reduce cross-talk between neighboring pixels and allow for a relatively simple planar device structure. However, with some minor modifications, the sensor concept can be transferred to the Si platform, which also provides the opportunity for low-cost monolithic integration of the electronic driver circuits, that are currently implemented off-chip. Our sensor consists of a two-dimensional array of $n = 14 \times 14 = 196$ pixels, each containing an MSM photodetector[28] that converts incident light into photocurrent. Each detector comprises interdigitated metal fingers on the SI-GaAs semiconductor. Photoexcited electrons and holes drift under an electrical field applied between the fingers, giving rise to an external current. The photoresponsivity of the device can be controlled by a bias voltage, as shown in Figure 2b, where the negative sign of the responsivity indicates a reversed current flow direction. The low background carrier concentration of the SI-GaAs wafer ($\sim 8 \times 10^6$ cm$^{-3}$) ensures full depletion of majority carriers. As a result, the electric field drops homogeneously in the space-charge region between the metal fingers, so that photogenerated carriers are efficiently swept out of the device. Low residual doping is also required to suppress dark current and reduce cross-talk between neighboring detectors. Finally, we verified an approximately linear illumination intensity-dependence of the photocurrent (Supplementary Figure S2), as required by equation (1).

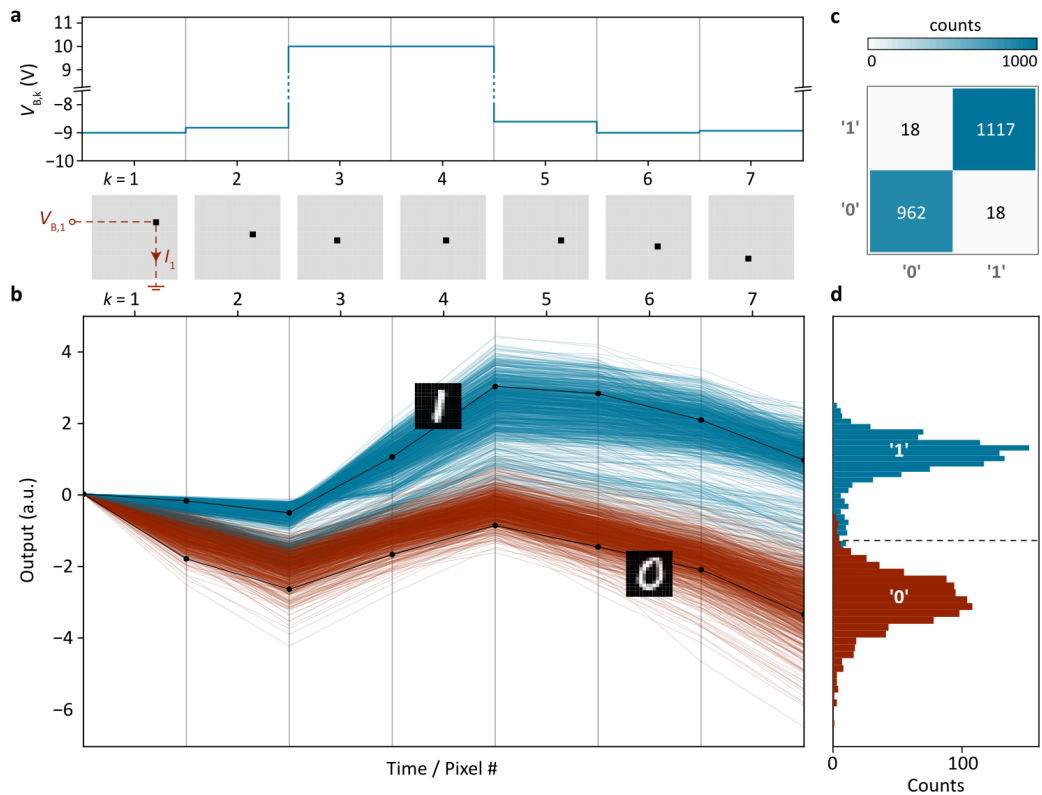

**Figure 3 | Image sensor operation and performance evaluation. a,** Relevant pixel locations $k$ (bottom) and applied bias voltages $V_{B,k}$ (top) for the binary classification task discussed in the main text. **b,** Temporal evolution of the sensor output for more than 2000 samples from the dataset. Red (blue) lines show cases in which a "0" ("1") has been projected onto the sensor. The black lines show two representative examples with corresponding MNIST digits. **c,** Experimental confusion matrix. A classification accuracy of 98.3% is achieved. **d,** Histogram of sensor output as determined from the measurements in b. The dashed line indicates the decision threshold.

As in CMOS sensor technology, detectors are addressed by row and column decoders. The readout is performed one pixel at a time, with relevant pixel locations $k$ and corresponding photoresponsivity values $R_k$ being determined from equation (2). Pixel-to-pixel variations and the nonlinear $R_k$-versus-$V_{B,k}$ behavior in Figure 2b are accounted for as discussed in Supplementary Figure S3. For details regarding the optical apparatus used for image projection, we refer to the Methods section.

We evaluated the sensor performance by the same binary classification task as discussed above. During a measurement, a bias voltage $V_{B,k}$ is applied to the $k$-th pixel via the row select line and the generated photocurrent is read out via the respective column line (Figure 3a). The resulting photocurrent is integrated over a period of time (~ms), before the next relevant detector is addressed by its row and column, and its output is added. We conducted this measurement for more than 2000 images of zeros and ones from the dataset. The bundle of curves, displayed in Figure 3b, shows the temporal evolution of the output for each of those samples. Traces in red show cases where a zero has been projected onto the chip; traces in blue correspond to a one. Two representative examples with corresponding images are shown as black lines. With each additional pixel measured, the red and blue traces separate further and the classification accuracy improves. At the end of a cycle (here, after 7 pixels), the output signal is compared to a threshold value and assigned to one of the two classes. Then the next cycle commences. Figure 3d shows a histogram of the sensor outputs after each cycle/sample, as determined from the measurements in Figure 3b. From the experimental confusion matrix, presented in Figure 3c, we determine a classification accuracy of 98.3%. The small deviation from the theoretical expectation (99.7%) and the shift of the decision threshold to below zero are attributed to device imperfections.

In summary, we presented a sensor that can be trained to classify images with an accuracy comparable to that of frame-based cameras by reading out the few most relevant pixels. We note that, although the use of tunable detector elements results in a particularly sleek and simplistic sensor design, the same principle can be transferred to conventional CMOS cameras with fixed pixel responsivities. The inner product $\mathbf{r}^T\mathbf{p}$ could there be realized, either in the analogue or digital domain, with additional electronics that is placed on the chip. We propose that such a system could be operated in a low-power mode, running the algorithm outlined above, and once a certain scene or gesture is detected, the system switches into a full-frame mode for further analysis.

# METHODS

**Sensor fabrication.** As a substrate we used a semi-insulating gallium arsenide (SI-GaAs) wafer, covered by 20 nm atomic layer deposition (ALD) grown $Al_2O_3$. A first metal layer was fabricated by evaporating Ti/Au (3/25 nm) through a mask created by electron-beam lithography (EBL). A 30-nm-thick $Al_2O_3$ oxide was then deposited using ALD, followed by a second lithography step and wet chemical etching in potassium hydroxide to define via-holes that connect the bottom and top metal layers and top metal with the GaAs substrate where necessary. Lastly, a top metal layer was added by another EBL process and Ti/Au (5/80 nm) evaporation. Prior to the metal evaporation, we removed the GaAs native oxide by a short dip of the sample in concentrated hydrochloric acid. We confirmed the continuity and solidity of the electrode structure by optical microscopy and electrical measurements in a wafer probe station. The sample was finally mounted in a chip carrier and wire-bonded.

**Optical setup.** A collimated, linearly polarized light beam (650 nm wavelength), produced by a semiconductor laser diode, illuminates a spatial light modulator (SLM, Hamamatsu), operated in intensity-modulation mode. On the SLM, the MNIST digits are displayed and the polarization of the light is rotated according to the pixel value. A polarizer with its optical axis oriented perpendicular to the polarization direction of the incident light acts as analyzer. The generated optical image is then projected onto the image sensor with a 20× long working distance objective (Mitutoyo). A schematic illustration of the apparatus is provided in Supplementary Figure S4.

**Data availability.** The data that support the findings of this study are available from the corresponding authors upon reasonable request.

**Acknowledgments:** We thank Max Andrews for providing a GaAs wafer. We acknowledge financial support from the Austrian Science Fund FWF (START Y 539-N16).

**Author contributions:** T.M. conceived the device concept. D.K.P., L.M. and D.K. fabricated the sensor. L.M. and T.M. programmed the machine learning algorithms. L.M. designed and built the experimental setup, carried out the measurements and analyzed the data. T.M. and L.M. prepared the manuscript. All authors discussed the results and commented on the manuscript.

**Competing financial interests:** The authors declare no competing financial interests.